\documentclass[trackchanges,twocolumn]{aastex7}
\usepackage{amsmath}
\usepackage{float}
\usepackage{graphicx}
\usepackage{booktabs}
\usepackage{bbold}
\usepackage{adjustbox}
\renewcommand\arraystretch{1.2}

\begin{document}

\title{General Implicit Runge-Kutta Integrators for Multifluid Gas-Dust Aerodynamic Drag}

\author{Giovanni Tedeschi-Prades}
\affiliation{University Observatory, Faculty of Physics, Ludwig-Maximilians-Universität München, Scheinerstr. 1, 81679 Munich, Germany}
\email[show]{giovanni.tedeschi@campus.lmu.de}  

\author{Til Birnstiel}
\affiliation{University Observatory, Faculty of Physics, Ludwig-Maximilians-Universität München, Scheinerstr. 1, 81679 Munich, Germany}
\affiliation{Exzellenzcluster ORIGINS, Boltzmannstr. 2, D-85748 Garching, Germany}
\email{til.birnstiel@lmu.de} 

\author{Klaus Dolag}
\affiliation{University Observatory, Faculty of Physics, Ludwig-Maximilians-Universität München, Scheinerstr. 1, 81679 Munich, Germany}
\affiliation{Max-Planck-Institut für Astrophysik, Karl-Schwarzschild-Straße 1, 85741 Garching, Germany}
\email{kdolag@mpa-garching.mpg.de} 

\author{Barbara Ercolano}
\affiliation{University Observatory, Faculty of Physics, Ludwig-Maximilians-Universität München, Scheinerstr. 1, 81679 Munich, Germany}
\affiliation{Exzellenzcluster ORIGINS, Boltzmannstr. 2, D-85748 Garching, Germany}
\email{ercolano@usm.uni-muenchen.de} 

\author{Mark Hutchison}
\affiliation{University Observatory, Faculty of Physics, Ludwig-Maximilians-Universität München, Scheinerstr. 1, 81679 Munich, Germany}
\affiliation{Hochschule für angewandte Wissenschaften München, Lothstraße 34, D-80335 München, Germany}
\email{gtedeschi@usm.lmu.de} 

\begin{abstract}
The integration of aerodynamic drag is a fundamental step in simulating dust dynamics in hydrodynamical simulations. We propose a novel integration scheme, designed to be compatible with Strang splitting techniques, which allows for the straightforward integration of external forces and hydrodynamic fluxes in general-purpose hydrodynamic simulation codes. Moreover, this solver leverages an analytical solution to the problem of drag acceleration, ensuring linear complexity even in cases with multiple dust grain sizes, as opposed to the cubic scaling of methods that require a matrix inversion step. This new General Implicit Runge-Kutta integrator (GIRK) is evaluated using standard benchmarks for dust dynamics such as DUSTYBOX, DUSTYWAVE, and DUSTYSHOCK. The results demonstrate not only the accuracy of the method but also the expected scalings in terms of accuracy, convergence to equilibrium, and execution time. GIRK can be easily implemented in hydrodynamical simulations alongside hydrodynamical steps and external forces, and is especially useful in simulations with a large number of dust grain sizes.
\end{abstract}


\section{Introduction}
Astrophysical dust is a fundamental component of a wide range of objects, from distant galaxies to nearby protoplanetary disks (PPDs). Its thermal emission allows us to probe astrophysical objects in the millimeter-infrared wavelength range, and its presence has an impact on both the thermodynamic and dynamical state of the surrounding gas. Dust grains scatter, absorb, and re-emit infrared radiation, controlling the temperature of the surrounding gas. From a dynamical point of view, dust continuously exchanges momentum with gas via aerodynamic drag, leading to a diverse array of gas-dust interactions that are often non-linear and difficult to predict without detailed numerical simulations. This is particularly true for larger grain sizes that are only partially coupled to the gas or in regions with high dust-to-gas ratios, both common scenarios in PPDs.

Dust dynamics has been the focus of many new developments in modern hydrodynamical simulation codes. In hybrid Eulerian-Lagrangian schemes, gas is evolved on a grid (either static or adaptive), while dust is simulated as a set of "superparticles" that move freely over the mesh \citep{Yang_2016,McKinnon_2018,Mignone_2019}. Other Eurlerian schemes, instead, include dust as an additional fluid alongside gas \citep{Benitez-Llambay_2019, Huang_2022}. In Lagrangian codes, dust has been treated either as a separate set of particles \citep{Monaghan1995,Laibe_2012} or via a `One-Fluid' approach, where each particle represents a gas–dust mixture rather than a single phase \citep{Laibe_2014}.

A key aspect of simulating dust dynamics, regardless of the numerical approach, is the integration of the drag term coupling the gas and dust. This drag term can be integrated either explicitly or implicitly; however, explicit integration requires the timestep of the simulation to be smaller than the shortest stopping time from the entire dust population. In many astrophysical systems, the stopping time for the most tightly coupled grains is much shorter than the simulation timestep, rendering explicit integration impractical. For this reason, implicit integration of aerodynamic drag is of paramount importance when simulating the dynamics of gas-dust mixtures.

Several methods have been developed to implicitly integrate the drag for a population of dust sizes, but these methods often require a matrix inversion to be carried out. While matrix inversion algorithms are readily available, these methods can be computationally expensive when a large number of dust sizes are needed, since the cost scales as $\mathcal{O}(N_d^3)$, where $N_d$ is the number of dust species. \cite{Krapp_2020} proposed a closed-form solution describing how velocity evolves under the influence of drag, assuming the momentum transfer between dust species is negligible. This solution only scales linearly with the number of dust grain sizes ($\mathcal{O}(N_d)$). 
\cite{Krapp_2024} combined this solution with the Diagonally Implicit Runge-Kutta (DIRK) integrator \citep{Ascher1997, Pareschi2005} to develop the first integration scheme that is second-order accurate with first-order complexity (called Multifluid DIRK, or MDIRK). While MDIRK offers significant computational advantages, its specialized Implicit-Explicit (IMEX) scheme for handling hydrodynamical fluxes and external forces limits its implementation to general-purpose hydrodynamic codes.

Our work builds upon the analytical solution presented in \cite{Krapp_2020} and the MDIRK integrator. In particular, we present a novel integration scheme, designed to be compatible with Strang splitting techniques, which allows for straightforward integration of external forces and hydrodynamical fluxes in general-purpose hydrodynamical codes. We introduce two versions of this integrator: one tailored to Strang splitting with a single hydrodynamic step, and another for the more commonly used two-step approach in second-order hydrodynamical simulations. This flexibility makes our scheme broadly applicable and easy to implement in a wide range of simulation frameworks.

In Section 2, we introduce the general formulation of the problem and briefly review the analytical solution derived by \cite{Krapp_2020}. We then develop a systematic procedure for translating desired properties, namely second-order accuracy and asymptotic stability, into a solvable set of algebraic conditions. We demonstrate that our new integration scheme satisfies all of these conditions simultaneously for a particular choice of parameters. In Section 3, we benchmark the scheme against a suite of test problems, confirming that it achieves all desired properties. Finally, Section 4 discusses the performance of the integrator and outlines potential applications.

\section{Methods}
In this paper, we consider a system consisting of gas and $N_\text{d}$ dust species that exchange momentum via aerodynamic drag. We restrict our discussion to a one-dimensional system; however, extensions to two and three dimensions are straightforward, since aerodynamic drag does not couple motions in perpendicular directions and can be treated independently. We denote gas properties with the subscript '$\text{g}$' and dust properties with the subscript '$\text{d}$'. Furthermore, an index $i=1, ..., N_\text{d}$ is used to label the different dust species. 

The continuity equations for the gas and the $N_\text{d}$ dust species read
\begin{equation}
    \begin{cases}
        \partial_t \rho_\text{g} + \partial_x(\rho_\text{g} v_\text{g}) = 0, \\[1mm]
        \partial_t \rho_{\text{d},i} + \partial_x(\rho_{\text{d},i} v_{\text{d},i}) = 0,
    \end{cases}
\end{equation}
where $\rho$ and $v$ denote the density and velocity, respectively. The momentum conservation equations are
\begin{equation}
    \begin{cases}
        \partial_t (\rho_\text{g} v_\text{g}) + \partial_x(\rho_\text{g} v_\text{g}^2) = -\partial_x P + G_\text{g} + \sum_{i=1}^{N_\text{d}} \rho_\text{g} \alpha_{\text{g},i} (v_{\text{d},i} - v_\text{g}),\\[1mm]
        \partial_t (\rho_{\text{d},i} v_{\text{d},i}) + \partial_x(\rho_{\text{d},i} v_{\text{d},i}^2) = G_{\text{d},i} - \rho_{\text{d},i} \alpha_{\text{d},i} (v_{\text{d},i} - v_\text{g}),
    \end{cases}
\end{equation}
where $P$ represents the gas internal pressure, $G$ any external force, and the remaining terms on the right-hand sides account for aerodynamic drag. Here, the momentum exchange rates $\alpha_{\text{d},i}$ are simply the inverses of the stopping times $t_{s,i}$. Enforcing momentum conservation, the gas momentum exchange rate $\alpha_{\text{g},i}$ is related to the dust term via
\begin{equation}
    \alpha_{\text{g},i} = \epsilon_i\alpha_{\text{d},i},
\end{equation}
where $\epsilon_i = \rho_\text{g} / \rho_{\text{d},i}$ is the dust fraction of the $i$-th dust specie. In the following, we will only refer to dust momentum exchange rates and drop the subscript 'd'.

In a more compact notation, the momentum conservation equations can be rewritten as
\begin{equation}
\label{eq:mom_cons_full}
    \partial_t\mathbf{u} + \partial_x\mathbf{\mathcal{F}} = \mathbf{M}\mathbf{u} + \mathbf{G},
\end{equation}
where:
\begin{equation}
    \mathbf{u} = (\rho_\text{g} v_\text{g},\; \rho_{\text{d},1}v_{\text{d},1},\; \dots,\; \rho_{\text{d},N_\text{d}}v_{\text{d},N_\text{d}}),
\end{equation}
\begin{equation}
    \mathbf{\mathcal{F}} = (\rho_\text{g} v_\text{g}^2 + P,\; \rho_{\text{d},1}v_{\text{d},1}^2,\; \dots,\; \rho_{\text{d},N_\text{d}}v_{\text{d},N_\text{d}}^2),
\end{equation}
\begin{equation}
    \mathbf{G} = (G_\text{g},\; G_{\text{d},1},\; \dots,\; G_{\text{d},N_\text{d}}),
\end{equation}
\begin{equation}
\mathbf{M} = \begin{pmatrix}
-\sum_{k=1}^{N_{\text{d}}} \epsilon_k \alpha_k & \alpha_1 & \alpha_2 & \cdots & \alpha_{N_\text{d}} \\
\epsilon_1 \alpha_1 & -\alpha_1 & 0 & \cdots & 0 \\
\epsilon_2 \alpha_2 & 0 & -\alpha_2 & \ddots & \vdots \\
\vdots & \vdots & \ddots & \ddots & 0 \\
\epsilon_{N_\text{d}} \alpha_{N_\text{d}} & 0 & \cdots & 0 & -\alpha_{N_\text{d}}
\end{pmatrix}.
\end{equation}
In what follows, we will first focus on numerical solutions to the homogeneous system
\begin{equation}
\label{eq:mom_cons}
    \partial_t\mathbf{u} = \mathbf{M}\mathbf{u},
\end{equation}
and later discuss how to incorporate the remaining terms from Eq.~\ref{eq:mom_cons_full}. 

\subsection{Explicit Integration}
The simplest approach is to integrate the drag term $\mathbf{M}\mathbf{u}$ explicitly. For example, using a first-order Euler method yields
\begin{equation}
\label{eq:impl_euler}
    \mathbf{u}^{n+1} = \mathbf{u}^n + \Delta t \, \mathbf{M}\mathbf{u}^n.
\end{equation}
The same explicit integration can also be carried out by higher-order schemes, such as Runge-Kutta integrators. However, explicit methods require the timestep to be restricted to the smallest stopping time in the simulation, often rendering the timestep prohibitively small \citep{Laibe_2014}.

\subsection{Implicit Integration}
Using an implicit Backward Euler scheme, Eq.~\ref{eq:mom_cons} can be integrated via matrix inversion:
\begin{equation}
\label{eq:impl_integration}
    \mathbf{u}^{n+1} = (1 - \Delta t\, \mathbf{M})^{-1}\mathbf{u}^n.
\end{equation}
Although this method removes the strict timestep restriction of the explicit scheme, it necessitates inverting a matrix, a process with a computational complexity of $\mathcal{O}(N_{\text{d}}^3)$.

For applications requiring a large number of dust bins, it is advantageous to avoid matrix inversion by employing more direct analytical solutions.

\subsection{Analytical Solution by \cite{Krapp_2020}}
To bypass matrix inversion, \cite{Krapp_2020} presented an analytical solution for the drag integration problem that reduces the complexity from third to first order, assuming that dust species exchange momentum only with the gas (and not among themselves). While their solution is formulated in terms of primitive variables, \cite{Krapp_2024} extended it to conserved variables and to the generalized problem of solving
\begin{equation}
\label{eq:general_impl_integration}
    (\mathbf{1} - \gamma\, \Delta t \, \mathbf{M})\mathbf{k} = \mathbf{M}\mathbf{q},
\end{equation}
for arbitrary $\mathbf{k}$ and $\mathbf{q}$ vectors, each having $N_\text{d}+1$ elements (one for each fluid), and with $\gamma$ being a free parameter. The solution for the gas component is then
\begin{equation}
    k_{\text{g}}^n = \frac{\mathcal{A} - q_{\text{g}}\,\mathcal{B}}{1 + \gamma\, \Delta t\, \mathcal{B}},
\end{equation}
and for the $i$-th dust component
\begin{equation}
    k_{d,i} = \frac{\alpha_i}{1 + \gamma\, \Delta t\, \alpha_i}\Bigl(\epsilon_i\, q_{\text{g}} - q_{{\text{d}},i} + \gamma\, \Delta t\, \epsilon_i\, k_\text{g}\Bigr),
\end{equation}
with the constants
\begin{equation}
\label{eq:anal_constants}
    \mathcal{A} = \sum_{k = 1}^{N_{\text{d}}} \frac{\alpha_k q_{\text{d},k}}{1 + \gamma\, \Delta t\, \alpha_{k}}, \quad  \mathcal{B} = \sum_{k=1}^{N_\text{d}} \frac{\epsilon_k\, \alpha_k}{1 + \gamma\, \Delta t\, \alpha_k}.
\end{equation}
This method solves Eq.~\ref{eq:mom_cons} without inverting a matrix, thereby reducing the computational complexity from $\mathcal{O}(N_\text{d}^3)$ to $\mathcal{O}(N_\text{d})$.

\subsection{Diagonally Implicit Runge-Kutta (DIRK) Integration from \cite{Krapp_2024}}
The analytical solution in Eqs. \ref{eq:general_impl_integration} - \ref{eq:anal_constants} can be incorporated into a second-order integrator scheme, such as the Diagonally Implicit Runge-Kutta (DIRK) integrator, as was done in \cite{Krapp_2024}. In this framework, the integration of Eq.~\ref{eq:mom_cons} is
\begin{equation}
\mathbf{u}^{n+1} = \mathbf{u}^n + \Delta t\, (1-\gamma)\, \mathbf{k}_1 + \Delta t \,\gamma\,\mathbf{k}_2,
\end{equation}
where $\gamma$ is a free parameter and the vectors $\mathbf{k}_1$ and $\mathbf{k}_2$ are determined from
\begin{equation}
\label{eq:DIRK_system}
\begin{cases}
    (1 - \gamma\, \Delta t\, \mathbf{M})\mathbf{k}_1 = \mathbf{M}\mathbf{u}_n,\\[1mm]
    (1 - \gamma\, \Delta t\, \mathbf{M})\mathbf{k}_2 = \mathbf{M}\Bigl(\mathbf{u}_n + (1-\gamma)\Delta t\, \mathbf{k}_1\Bigr).
\end{cases}
\end{equation}
Eq. \ref{eq:DIRK_system} presents two equations, both with the same structure as Eq. \ref{eq:general_impl_integration}. The analytical solution can thus be used to find a solution for both $\mathbf{k}_1$ and $\mathbf{k}_2$. This integrator is termed “diagonally implicit” because $\mathbf{k}_1$ can be computed independently of $\mathbf{k}_2$, allowing sequential computation rather than solving a coupled system. \cite{Krapp_2024} coupled this integrator with an IMEX scheme for external forces and hydrodynamical fluxes, and named the combined method MDIRK. The explicit forms of $\mathbf{k}_1$ and $\mathbf{k}_2$ are provided in \citealp[(Appendix A)]{Krapp_2024}.

\subsection{Imposing Desirable Properties}\label{sec:impose_properties}
Before introducing our new method, we reframe the discussion into the more general problem of solving Eq.~\ref{eq:mom_cons_full}, which includes external forces and hydrodynamical fluxes. In numerical simulations, particularly when implicit integration is applied to source terms (such as aerodynamic drag), operator splitting is typically used. Popular choices include Strang splitting (a second-order method) or Lie-Trotter splitting (a first-order method where operators are applied for a full step consecutively). Other possible solutions are mixed IMEX methods, which require more specific time-stepping schemes, like the MDIRK integrator \citep{Krapp_2024}.

Ideally, we want the full integration scheme used to solve Eq.~\ref{eq:mom_cons_full} to preserve the second-order accuracy of the drag integrator. Additionally, we aim to enforce second-order convergence in the presence of external forces, a property that is not guaranteed in general. This was already discussed in Appendix C of \cite{Krapp_2024}, where the authors describe how their DIRK integrator, coupled with a Strang splitting scheme, fails to both achieve second-order accuracy and to converge to the correct equilibrium in the presence of external forces.

In the following, we introduce a framework for assessing second-order convergence and accuracy in a given integration scheme and show how any DIRK integrator, when combined with a Strang splitting scheme, fails to satisfy both simultaneously.

We begin by defining the hydrodynamical step operator, $\mathcal{H}_{\Delta t}$, which applies an explicit integration of the hydrodynamical fluxes and external force terms to the state vector $\mathbf{u}$. Since our focus here is on the convergence properties of the scheme in the presence of a constant external force, we restrict $\mathcal{H}_{\Delta t}$ to the simplified form
\begin{equation}
    \mathcal{H}_{\Delta t}\mathbf{u} = \mathbf{u} + \mathbf{G_0} \Delta t,
\end{equation}
where $\mathbf{G_0}$ is a constant external force.

Similarly, we define the drag operator $\mathcal{D}_{\Delta t}$, whose action on the state vector $\mathbf{u}$ depends on the chosen implicit integrator. For instance, the DIRK integrator \citep{Krapp_2024} can be recast as
\begin{equation}
    \mathcal{D}_{\Delta t}\mathbf{u} = \Bigl(1 + \mathbf{A}\Delta t + \gamma(1-\gamma)\mathbf{A}^2 \Delta t^2\Bigr)\mathbf{u},
\end{equation}
where $\mathbf{A} = (1-\gamma\,\Delta t\,\mathbf{M})^{-1}\mathbf{M}$.

These two operators can be combined using a Strang splitting scheme, such as 
\begin{equation}
\label{eq:strang_splitting}
    \mathbf{u}^{n+1} = \mathcal{D}_{\Delta t/2}\,\mathcal{H}_{\Delta t}\,\mathcal{D}_{\Delta t/2}\,\mathbf{u}^n.
\end{equation}
To facilitate analysis, we decouple the system by diagonalizing the matrix $\mathbf{M}$. Let $\mathbf{P}$ be the transformation matrix that diagonalizes $\mathbf{M}$. Applying $\mathbf{P}^{-1}$ to Eq. \ref{eq:strang_splitting} yields
\begin{equation}
\label{eq:diagonalized_equations}
    \hat{u}^{n+1}_k = \left(\mathbf{P}^{-1}\,\mathcal{D}_{\Delta t/2}\,\mathcal{H}_{\Delta t}\,\mathcal{D}_{\Delta t/2}\,\mathbf{u}^n\right)_k,
\end{equation}
where $\hat{u}^{n+1}_k$ is the $k$-th entry of the eigenspace-projected vector $\mathbf{\hat{u}}^{n+1} = \mathbf{P}^{-1}\mathbf{u}^{n+1}$, and corresponds to the eigenvalue $\lambda_k$ of $\mathbf{M}$. The zeroth component of $\mathbf{\hat{u}}^{n+1}$ corresponds to the bulk motion of the gas-dust mixture. All other components correspond to the velocity differences between gas and individual dust species, and are the focus of our analysis.

With the system decoupled, we can expand Eq. \ref{eq:diagonalized_equations} in series and impose convergence and accuracy conditions. Importantly, we will need to perform the expansions both for $\Delta t \to 0$ and $\Delta t \to \infty$, to verify consistency with the differential equation’s exact solution and ensure asymptotic stability in the stiff regime, respectively.

In the limit $\Delta t \to 0$, the expansion of the numerical update in Eq. \ref{eq:diagonalized_equations} will contain terms proportional to $\hat{u}_k^n$ as well as terms proportional to the projected external force $\hat{G}_k$
\begin{equation}
\label{eq:taylor_numerical}
\begin{split}
    \hat{u}_k^{n+1} &\simeq \hat{u}_k^n[1 + A_1\Delta t + A_2\Delta t^2 + \mathcal{O}(\Delta t^3)]+\\
    &+ \hat{G}_k[B_1\Delta t + B_2\Delta t^2 + \mathcal{O}(\Delta t^3)].
\end{split}
\end{equation}
To impose second-order convergence, we need to compare this against the exact solution of the diagonalized version of Eq. \ref{eq:mom_cons_full}
\begin{equation}
    \frac{d\hat{u}_k}{dt} = \lambda_k\hat{u}_k + \hat{G}_k.
\end{equation}
Expanding it in Taylor series, we obtain the following expected solution for $\Delta t \to 0$
\begin{equation}
\label{eq:taylor_exact}
    \begin{split}
        \hat{u}_k(t+\Delta t) &\simeq \hat{u}_k(t)[1 + \lambda_k \Delta t + \frac{1}{2}\lambda_k^2\Delta t^2 + \mathcal{O}(\Delta t^3)] + \\
        & + \hat{G}_k[\Delta t + \frac{1}{2}\lambda_k\Delta t^2 + \mathcal{O}(\Delta t^3)].
    \end{split}
\end{equation}
Comparing Eqs. \ref{eq:taylor_numerical} and \ref{eq:taylor_exact} we obtain constraints on $A_1$, $A_2$, $B_1$, and $B_2$ that ensure second-order convergence for $\Delta t \to 0$. Higher-order schemes can be constructed by matching higher-order terms in the expansion.

For the opposite limit, $\Delta t \to \infty$, the process is similar. We first expand the numerical update in Eq. \ref{eq:diagonalized_equations}
\begin{equation}
\label{eq:series_numerical}
\begin{split}
    \hat{u}_k^{n+1} &\simeq \hat{u}_k^n[C_0 + C_1\Delta t^{-1} + \mathcal{O}(\Delta t^{-2})]+\\
    &+ \hat{G}_k[D_0 + D_1\Delta t^{-1} + \mathcal{O}(\Delta t^{-2})].
\end{split}
\end{equation}
In the absence of external forces, the expected behavior for $k \neq 0$ is $\hat{u}_k^{n+1} \to 0$  as $\Delta t \to \infty$, since the velocity difference between gas and dust is damped out. Hence, for second-order convergence of the drag terms, we require $C_0 = 0$ and $C_1 = 0$, so that the leading term in the error is of second order. With external forces, the equilibrium solution is $\hat{u}_k^{n+1} \to \hat{G}_K/|\lambda_k|$ \citep{Krapp_2024}, which constrains the value of $D_0$. To ensure second-order convergence to this equilibrium, we additionally require $D_1=0$.

In practice, it is often not possible to find a single set of parameters that satisfies all constraints in both limits. However, we can switch between two parameter sets depending on whether the timestep is smaller or larger than the maximum stopping time $t_\text{s}^{\text{max}}$. For instance, the DIRK integrator, coupled with the Strang splitting scheme in Eq. \ref{eq:strang_splitting}, is able to satisfy nearly all our desired properties with the following choices of $\gamma$
\begin{equation}
\gamma = 
\begin{cases}
    1 \pm \displaystyle\frac{1}{\sqrt{2}} \quad \Delta t < t_\text{s}^{\text{max}},\\
    2 \pm \sqrt{2} \quad \Delta t > t_\text{s}^{\text{max}}.
\end{cases}
\end{equation}
The only unsatisfied condition is second-order convergence to the equilibrium solution in the presence of external forces (i.e., $D_1 \neq 0$). To address this, we must extend the approach of \cite{Krapp_2024} to a more general, non-diagonal, implicit Runge–Kutta integrator.

\subsection{General Implicit Runge-Kutta (GIRK) Integrator}
Since the DIRK integrator cannot satisfy all the desired properties simultaneously, we extend our approach to a more General Implicit Runge-Kutta (GIRK) integrator. Starting from the backbone of the Runge-Kutta integration scheme, new integrators can be constructed by constraining the free parameters in the Butcher's Table according, for instance, to the desired integration order or the stability properties or the performances for a specific problem desired. While some general methods have been developed to generate new Runge-Kutta integrators \citep{Maurya2021,Mizerova2024}, we will here derive our own set of parameters that can satisfy all the properties introduced in Section \ref{sec:impose_properties}.

The update step of a GIRK is of the form
\begin{equation}
        \mathbf{u}^{n+1} = \mathbf{u}^n + \Delta t b \mathbf{k}_1 + \Delta t (1-b)\mathbf{k}_2.
\end{equation}
Here, $\mathbf{k}_1$ and $\mathbf{k}_2$ are obtained by solving the coupled system
\begin{equation}
\begin{cases}
    (1 - \gamma_1 \Delta t\, \mathbf{M})\,\mathbf{k}_1 = \mathbf{M}\mathbf{u}_n + \Delta t\, \beta_1\, \mathbf{M}\,\mathbf{k}_2,\\[1mm]
    (1 - \gamma_2 \Delta t\, \mathbf{M})\,\mathbf{k}_2 = \mathbf{M}\mathbf{u}_n + \Delta t\, \beta_2\, \mathbf{M}\,\mathbf{k}_1.
\end{cases}
\end{equation}
This scheme introduces five parameters: $\gamma_1$, $\gamma_2$, $\beta_1$, $\beta_2$, and $b$. Although the resulting expressions are more involved than in the DIRK case, they still admit an analytical solution. The gas components are given by
\begin{equation}
\label{eq:sol_beg}
    k_{1\text{g}} = \frac{\beta_1 \Delta t\, C_1 (A_2 - B_2 u_\text{g}) - D_2 (A_1 - B_1 u_\text{g})}{\beta_1\beta_2\Delta t^2 C_1 C_2 - D_1 D_2},
\end{equation}
\begin{equation}
    k_{2\text{g}} = \frac{\beta_2 \Delta t\, C_2 (A_1 - B_1 u_\text{g}) - D_1 (A_2 - B_2 u_\text{g})}{\beta_1\beta_2\Delta t^2 C_1 C_2 - D_1 D_2},
\end{equation}
and the dust components for the $i$-th specie are
\begin{equation}
\begin{split}
    &k_{1\text{d},i} = \alpha_i\,\Lambda_i\Bigl[(u_\text{g}\epsilon_i - u_{\text{d},i})\Bigl(1+\alpha_i\Delta t(\gamma_2-\beta_1)\Bigr)\\[1mm]
    &\quad\quad +\, k_{1\text{g}}\,\epsilon_i\Delta t\Bigl(\gamma_1 + \alpha_i\Delta t\,(\gamma_1\gamma_2 - \beta_1 \beta_2)\Bigr) + k_{2\text{g}}\,\beta_1 \epsilon_i\Delta t\Bigr],
\end{split}
\end{equation}
\begin{equation}
\label{eq:sol_end}
\begin{split}
    &k_{2\text{d},i} = \alpha_i\,\Lambda_i\Bigl[(u_\text{g}\epsilon_i - u_{\text{d},i})\Bigl(1+\alpha_i\Delta t(\gamma_1-\beta_2)\Bigr)\\[1mm]
    &\quad\quad +\, k_{2\text{g}}\,\epsilon_i\Delta t\Bigl(\gamma_2 + \alpha_i\Delta t\,(\gamma_1\gamma_2 -\beta_1\beta_2)\Bigr) + k_{1\text{g}}\,\beta_2 \epsilon_i\Delta t\Bigr].
\end{split}
\end{equation}
All the constants used in these equations are defined in Appendix \ref{sec:appendix_A}. Despite the added complexity, this GIRK integrator remains linear in computational cost, as it avoids matrix inversion and instead relies on the analytical expressions derived in \cite{Krapp_2024}. The additional parameters also introduce more degrees of freedom, which can be used to enforce the full set of desirable properties discussed earlier. The drag operator associated with the GIRK operator can be written as
\begin{equation}
\begin{split}
\label{eq:Drag_GIRK_DHD}
    &\mathcal{D}_{\Delta t}\mathbf{u} = \Bigl(1 - \Delta t^2\,\beta_1\beta_2\,\mathbf{A}_1\,\mathbf{A}_2\Bigr)^{-1} \Bigl(1 + b\,\Delta t\,\mathbf{A}_1\\[1mm]
    &+ (1 - b)\,\Delta t\,\mathbf{A}_2 + \Bigl[\beta_2(1 - b) + b\beta_1 - \beta_1\beta_2\Bigr]\Delta t^2\,\mathbf{A}_1\,\mathbf{A}_2\Bigr),
\end{split}
\end{equation}
where $\mathbf{A}_1 = (1-\gamma_1\Delta t\,\mathbf{M})^{-1}\mathbf{M}$ and similarly for $\mathbf{A}_2$.

\subsection{Strang-Splitting Schemes and Parameter Values}\label{sec:parameters_GIRK}
We now propose two Strang splitting schemes to couple GIRK within hydrodynamical simulation codes. The first, simpler scheme is the one already described as
\begin{equation}
    \mathbf{u}^{n+1} = \mathcal{D}_{\Delta t/2}\,\mathcal{H}_{\Delta t}\,\mathcal{D}_{\Delta t/2}\,\mathbf{u}^n.
\end{equation}

Having more free parameters, compared to the DIRK integrator, we are able to enforce additional properties. For instance, expanding it for $\Delta t \to 0$, we find that third-order convergence can be achieved with the following parameter choices\footnote{The calculations were carried out using a \texttt{Mathematica} notebook that can be downloaded at the following link: \url{https://github.com/Giovanni-Tedeschi/GIRK_MathematicaNotebook}}

\begin{equation}
\label{eq:params_DHD_small}
    \begin{cases}
        b = 1,\\[1mm]
        \beta_1 = \displaystyle\frac{1}{2} - \gamma_1,\\[2mm]
        \beta_2 = \displaystyle\frac{1 - 3\gamma_1 - 3\gamma_2 + 6\gamma_1\gamma_2}{3-6\gamma_1},
    \end{cases}
    \quad \text{if } \Delta t <  t_\text{s}^{\text{max}}.
\end{equation}
The values of $\gamma_1$ and $\gamma_2$ are still unconstrained, and can be tuned to better performances by minimizing the errors in benchmark tests. The values we suggest after some testing are $\gamma_1 = 1$ and $\gamma_2 = 0$.

A similar procedure in the opposite limit leads to:
\begin{equation}
    \begin{cases}
        b = 0,\\[1mm]
        \gamma_2 = 2 - \gamma_1,\\[1mm]
        \beta_1 = \displaystyle\frac{2 - 2 \gamma_1 + \gamma_1^2}{2 - \gamma_1},\\[1mm]
        \beta_2 = \gamma_1 - 2,
    \end{cases}
    \quad \text{if } \Delta t > t_\text{s}^{\text{max}},
\end{equation}
where the $\gamma_1$ is, again, unconstrained. We propose, after some testing, $\gamma_1 = 1.0$.

The second Strang splitting scheme we propose is designed to include two hydrodynamical steps per update, as this is usually the case in most second-order hydrodynamical simulation codes. One possibility to derive such an algorithm is to chain the previous scheme twice
\begin{equation}
        \mathbf{u}^{n+1} = \mathcal{D}_{\Delta t/4}\,\mathcal{H}_{\Delta t/2}\,\mathcal{D}_{\Delta t/4}\,\mathcal{D}_{\Delta t/4}\,\mathcal{H}_{\Delta t/2}\,\mathcal{D}_{\Delta t/4}\,\mathbf{u}^n.
\end{equation}
Alternatively, we can optimize this splitting scheme by combining the two central $\mathcal{D}_{\Delta t/4}$ operators
\begin{equation}
\label{eq:strang_splitting_2}
        \mathbf{u}^{n+1} = \mathcal{D}_{\Delta t/4}\,\mathcal{H}_{\Delta t/2}\,\mathcal{D}_{\Delta t/2}\,\mathcal{H}_{\Delta t/2}\,\mathcal{D}_{\Delta t/4}\,\mathbf{u}^n.
\end{equation}
This new splitting scheme has the same convergence and accuracy scalings for $\Delta t \to 0$ as the previous one, and can be used with the same parameters as in Eq. \ref{eq:params_DHD_small}. In the stiff limit, however, third-order convergence can be achieved with the following parameters
\begin{equation}
\begin{cases}
    b = 1,\\[1mm]
    \gamma_2 = 3 - \gamma_1,\\[1mm]
    \beta_1 = -\gamma_1 - 1,\\[1mm]
    \beta_2 = \displaystyle\frac{\gamma_1^2 - 3 \gamma_1 + 4}{1 + \gamma_1},
\end{cases}
\quad \text{if } \Delta t > t_\text{s}^{\text{max}},
\end{equation}
with $\gamma_1 = 1$. Table~\ref{tab:params} summarizes the recommended parameter values for both splitting schemes and both timestep regimes.
\begin{table}[h]
\centering
\caption{\label{tab:params} Parameters of the two GIRK integrators for both Strang splitting schemes.}
\begin{tabular}{c|c|c|c|c}
& \multicolumn{2}{c|}{$\mathcal{D}_{\Delta t/2}\,\mathcal{H}_{\Delta t}\,\mathcal{D}_{\Delta t/2}$} & \multicolumn{2}{c}{$\mathcal{D}_{\Delta t/4}\,\mathcal{H}_{\Delta t/2}\,\mathcal{D}_{\Delta t/2}\,\mathcal{H}_{\Delta t/2}\,\mathcal{D}_{\Delta t/4}$} \\  
& $\Delta t < t_{s}^{\text{max}}$ & $\Delta t > t_{s}^{\text{max}}$ & $\Delta t < t_{s}^{\text{max}}$ & $\Delta t > t_{s}^{\text{max}}$ \\
\midrule
$\gamma_1$ & 1 & 1.0 & 1 & 1.0\\
$\gamma_2$ & 0 & 1.0 & 0 & 2.0\\
$\beta_1$ & -1/2 & 1.0 & -1/2 & -2.0\\
$\beta_2$ & 2/3 & -1.0 & 2/3 & 1.0\\
$b$ & 1.0 & 0.0 & 1.0 & 1.0\\
\bottomrule
\end{tabular}
\end{table}

\section{Results}
To test our newly developed GIRK implicit integrators and compare them against the MDIRK integrator, we created a one-dimensional Riemann solver code for gas–dust mixtures called \texttt{pigpen}\footnote{\texttt{pigpen} can be freely downloaded at \dataset[doi:10.5281/zenodo.17329227]{https://doi.org/10.5281/zenodo.17329227}. In the same repository, all the tests performed in this section can be found as well as a \texttt{jupyter} notebook to reproduce the plots}. This is a lightweight and flexible codebase, particularly well-suited for testing the different implicit integrators discussed above. The code supports both an exact Riemann solver and an HLL Riemann solver. However, only the exact solver is used in the tests presented here. Additionally, we employ a simple linear piecewise reconstruction method with a van Leer slope limiter \citep{VanLeer1977} to approximate the inter-cell values, before computing the fluxes between neighboring cells. The timestep is selected using a simple CFL condition
\begin{equation}
\label{eq:CFL}
    \Delta t_{\text{CLF}} = C_{\text{CFL}}\frac{\Delta x}{v_{\text{sig}}^{\text{max}}},
\end{equation}
where $C_{CFL} = 0.1$, $\Delta x$ is the (constant) spacing between the cells, and the maximum signal velocity is computed as
\begin{equation}
    v_{\text{sig}}^{\text{max}} = \max_n\left(c_\text{s}^n + v_\text{g}^n + \sum_{i=1}^{N_\text{d}}v_{\text{d},i}^n\right) \qquad n=1,...,N_{\text{cells}}
\end{equation}
where $c_\text{s}$ is the sound speed and the $n$ index spans over all cells. All the GIRK implicit integrators described in the previous section were implemented in \texttt{pigpen}, as well as the MDIRK integrator by \cite{Krapp_2024}. These methods will be tested and compared across a suite of standard multi-fluid dust dynamics benchmarks.

\subsection{DUSTYBOX}
\begin{figure*}[ht!]
    \centering
    \plotone{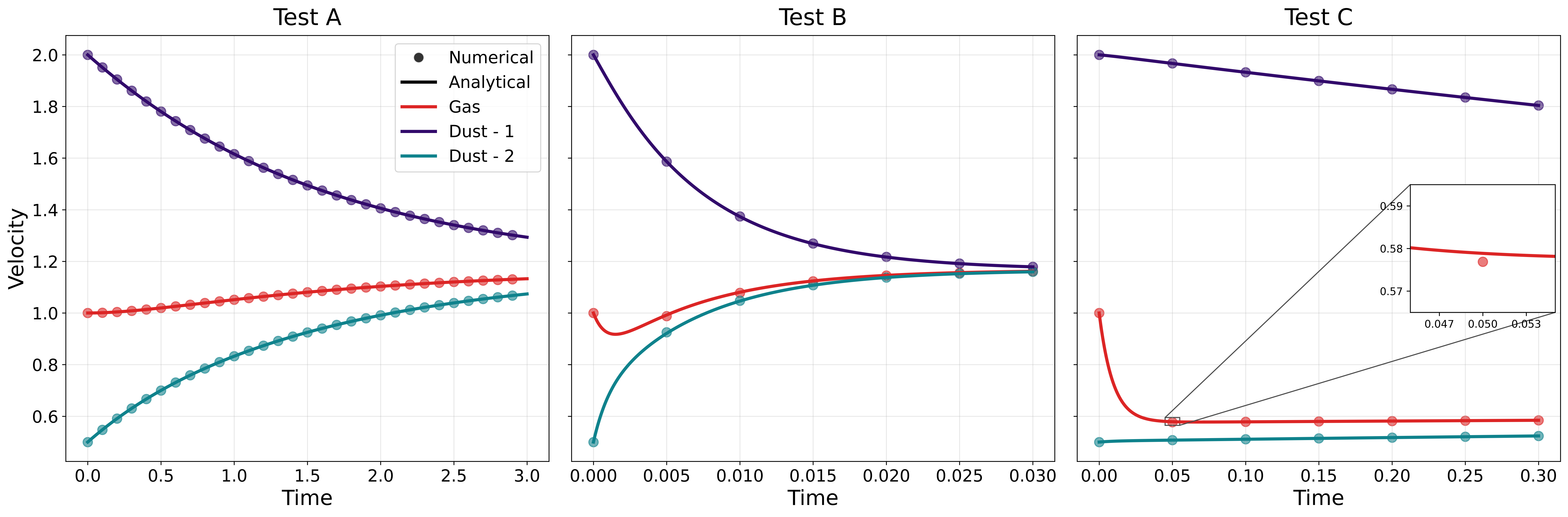}
    \caption{Simulation outputs and analytical solutions for the three DUSTYBOX benchmark tests described in \cite{Huang_2022}. The system is composed of a gas fluid and two dust fluids. The simulation outputs are shown as dots, while the analytical solution is drawn as a continuous line. For the non-stiff test A, the timestep is chosen according to the CFL condition Eq. \ref{eq:CFL}, while for the stiff tests B and C, the timestep is, respectively, $\Delta t = 0.005$ and $\Delta t = 0.05$. The integrator used (GIRK with $\mathcal{D}_{\Delta t/2}\mathcal{H}_{\Delta t}\mathcal{D}_{\Delta t/2}$ Strang splitting) is able to correctly capture the evolution of the mixture for all stiffness regimes. For Test C, we also show how much the integrator undershoots the analytical solution for the gas velocity, before converging.}
    \label{fig:DUSTYBOX}
\end{figure*}

In this standard test for dust dynamics, the DUSTYBOX test, we replicate the one-dimensional collision experiment proposed by \cite{Huang_2022}. Gas and two distinct dust populations are initialized with different velocities and densities. Under the influence of aerodynamic drag alone, the system evolves toward an equilibrium state at the center-of-mass velocity. The analytical solution to this problem is given by
\begin{equation}
    v(t) = v_{\text{COM}} + c_1 e^{\lambda_1 t} + c_2 e^{\lambda_2 t},
\end{equation}
where $v_{\text{COM}}$ is the center-of-mass velocity, and the parameters $c_1$, $c_2$, $\lambda_1$, and $\lambda_2$ are provided in \citealp[Table 1]{Huang_2022}, along with the corresponding initial conditions. We consider three setups: a non-stiff case (Test A), a stiff case with large drag (Test B), and a stiff case with a large dust-to-gas ratio (Test C). Figure~\ref{fig:DUSTYBOX} shows the simulation results compared to the analytical solution.

While for Test A we choose the timestep according to the CFL condition in Eq. \ref{eq:CFL}, for Tests B and C we fix it to, respectively, $\Delta t = 0.005$ and $\Delta t = 0.05$. This choice follows \citet{Krapp_2024} and is intended to investigate the monotonic convergence of the integrator. Since the GIRK integrator is not fully implicit, it may introduce small oscillations around the expected solution in stiff regimes, thereby preventing strictly monotonic convergence. Figure~\ref{fig:DUSTYBOX} shows, for Test C, a zoom-in of the gas velocity where a small oscillation is indeed visible before the solution converges. However, the amplitude of this oscillation is significantly smaller than that reported in Figure~2 of \citet{Krapp_2024}. 

\begin{figure}[ht!]
    \centering
    \includegraphics[width=0.45\textwidth]{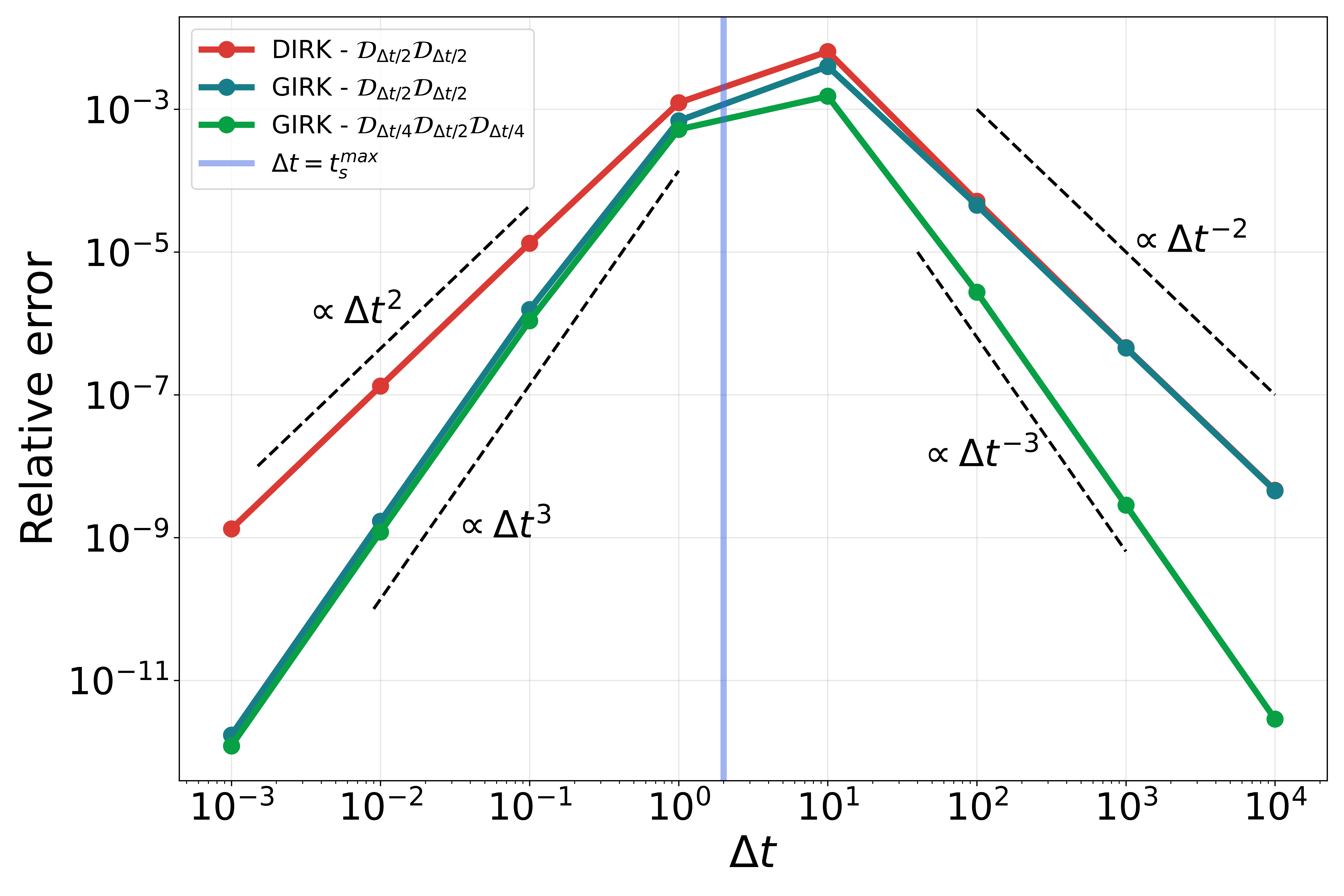}
    \caption{Convergence orders for the DIRK and the GIRK integrations for different Strang operator splitting schemes on Test A of \cite{Huang_2022}. The error is computed using Eq. \ref{eq:rel_error}. All integrators follow their expected scalings, given the properties imposed as in Section \ref{sec:impose_properties} and \ref{sec:parameters_GIRK}.}
    \label{fig:scaling}
\end{figure}

To quantify accuracy, we compute the relative error of the numerical solution as
\begin{equation}
\label{eq:rel_error}
    e = \frac{1}{N_{\text{step}}} \sum_{k=1}^{N_{\text{step}}}\sum_{i=0}^2\frac{|v_i(t_k) - v_i^{\text{num}}(t_k)|}{v_i(t_k)},
\end{equation}
where $N_{\text{step}}$ is the number of timesteps, $t_k$ denotes the time at step $k$, and the index $i$ refers to either the gas phase or one of the two dust phases. By fixing the simulation timestep, we evaluate the relative error for various integration methods on Test A. Figure~\ref{fig:scaling} compares the DIRK (with two drag integration steps) and the GIRK methods (with both two and three drag integration steps, corresponding to the Strang splitting schemes in Eq. \ref{eq:strang_splitting} and \ref{eq:strang_splitting_2}). As shown, the DIRK method achieves second-order accuracy for both large and small timesteps, while the GIRK methods can reach third-order accuracy in some regimes. This behaviour reflects the conditions that we imposed on the GIRK methods in Section \ref{sec:parameters_GIRK}.

\subsection{External Force Damping}
\begin{figure}[h!]
    \centering
    \plotone{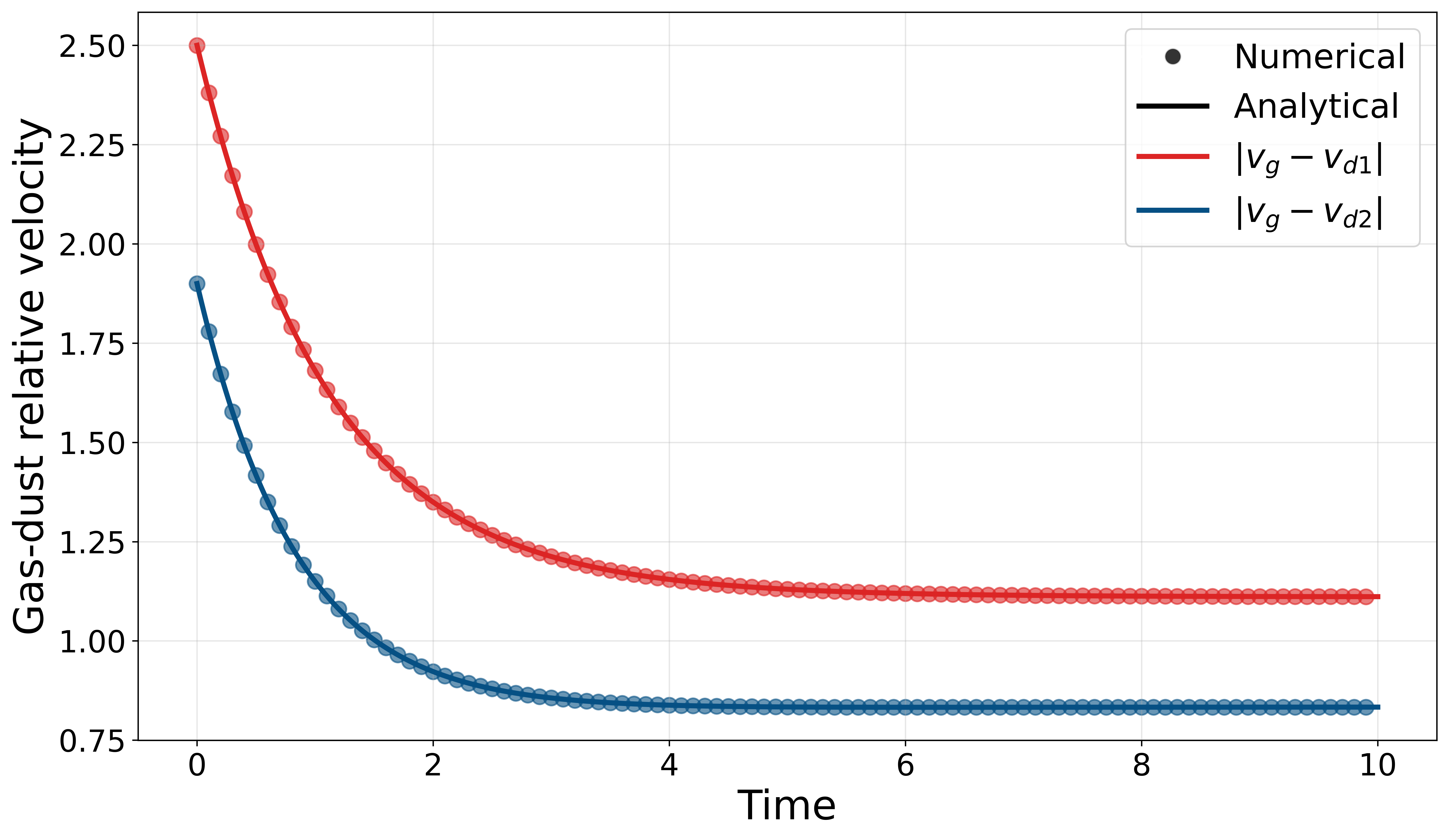}
    \caption{External force damping benchmark test. The dots represent the simulation output, while the solid line shows the analytical solution. The simulation was carried out using the GIRK drag integrator with the $\mathcal{D}_{\Delta t/2}\mathcal{H}_{\Delta t}\mathcal{D}_{\Delta t/2}$ Strang splitting scheme. As expected, the simulation follows the expected solution and converges to the correct asymptotic value.}
    \label{fig:ext_force}
\end{figure}

\begin{figure}[ht!]
    \centering
    \plotone{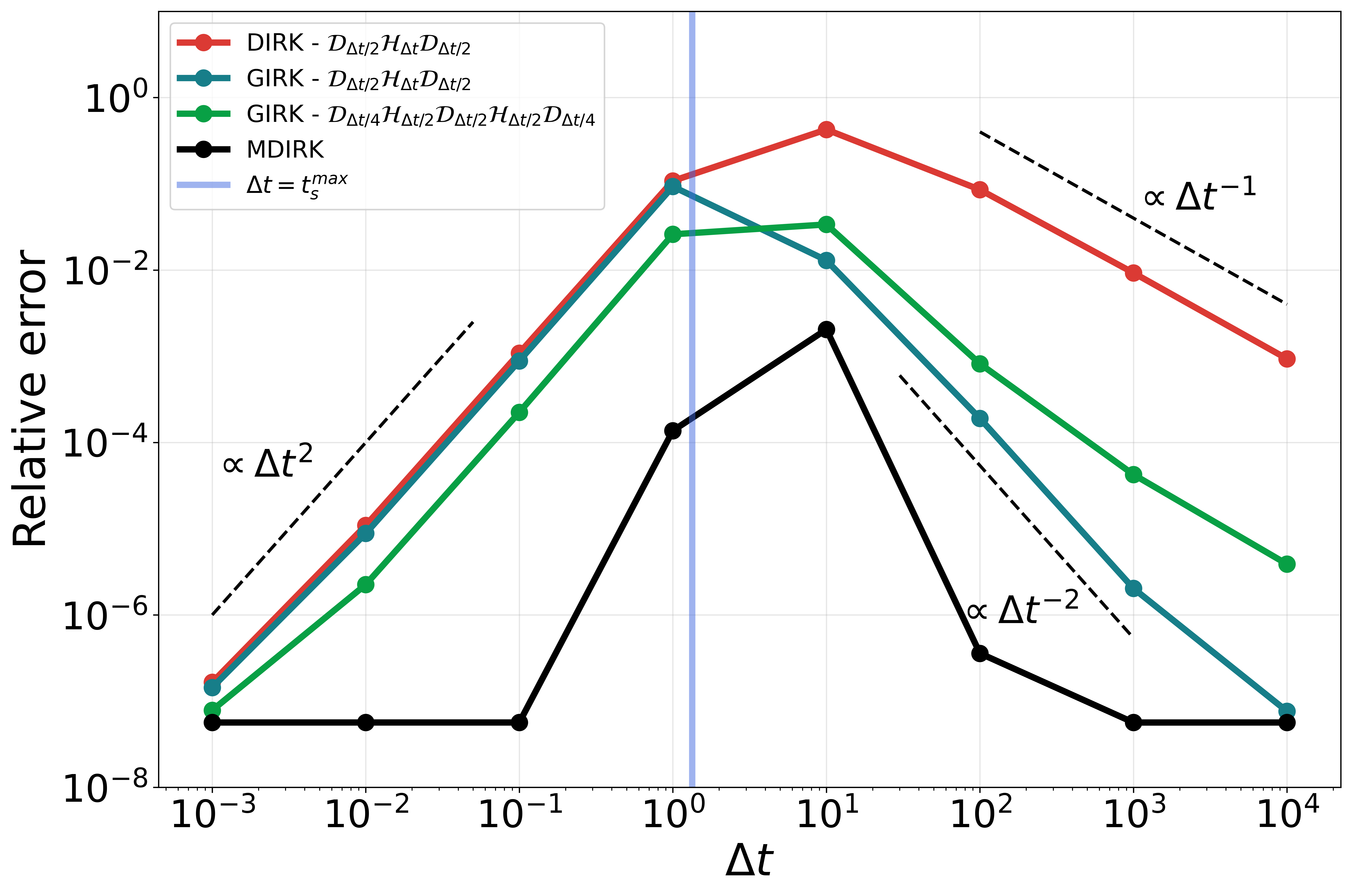}
    \caption{Relative error of the final simulation output compared to the analytical asymptotic solution, computed using Eq.~\ref{eq:rel_error}. The MDIRK method performs best, while the DIRK integrator coupled with Strang splitting converges only at first order in the stiff regime. The GIRK integrators, however, show significantly improved accuracy and faster convergence across both splitting schemes.}
    \label{fig:convergence}
\end{figure}

In the next test, we introduce an external force to evaluate the convergence behavior of the integration schemes in the presence of additional source terms. We adopt the benchmark described in Section 3.2 of \cite{Krapp_2024}, which provides both the analytical solution and initial conditions. In Figure \ref{fig:ext_force} we show the simulation output for the GIRK integrator in the $\mathcal{D}_{\Delta t/2}\mathcal{H}_{\Delta t}\mathcal{D}_{\Delta t/2}$ splitting scheme. As shown, the numerical solution correctly converges toward the expected asymptotic state.
To assess convergence quantitatively, Figure~\ref{fig:convergence} shows the relative error of the final simulation output compared to the analytical asymptotic solution, for various timestep values and integration schemes. Among the tested methods, the MDIRK integrator from \cite{Krapp_2024} delivers the best overall performance. However, it is also the least flexible for general implementation within hydrodynamical codes. The DIRK integrator, when coupled with a Strang splitting scheme, shows, as expected, only first-order convergence for large values of $\Delta t$, leading to substantial relative errors when the timestep approaches $t_{\text{s}}^{\text{max}}$. In contrast, the GIRK integrators yield significantly better agreement with the analytical solution and converge more rapidly. While the $\mathcal{D}_{\Delta t/2}\mathcal{H}_{\Delta t}\mathcal{D}_{\Delta t/2}$ splitting scheme delivers second order convergence for $\Delta t > t_{\text{s}}^\text{max}$, we notice that the $\mathcal{D}_{\Delta t/4}\mathcal{H}_{\Delta t}\mathcal{D}_{\Delta t/2}\mathcal{H}_{\Delta t}\mathcal{D}_{\Delta t/4}$ offers only a mixed first/second-order convergence.

\subsection{DUSTYWAVE and Complexity Scaling}
\begin{figure*}[ht!]
    \centering
    \includegraphics[width=1.0\linewidth]{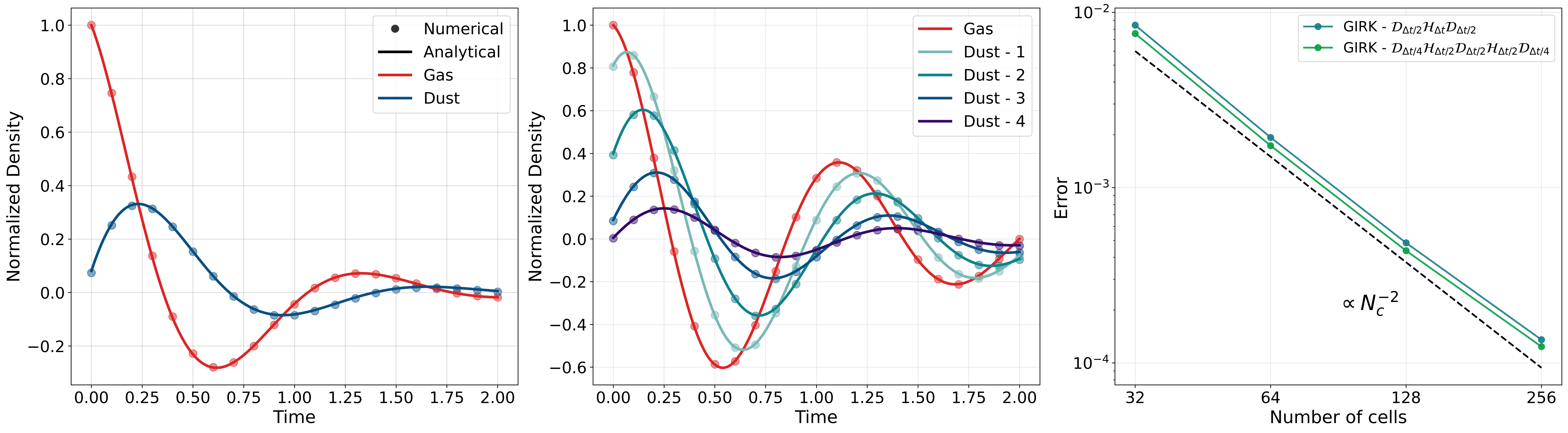}
    \caption{Simulation outputs and analytical solution of the DUSTYWAVE benchmark test for both a single dust specie (left panel) and four dust species (central panel). The plots show the time evolution of the normalized density perturbation. The simulation was carried out using the $\mathcal{D}_{\Delta t/2}\,\mathcal{H}_{\Delta t}\,\mathcal{D}_{\Delta t/2}$ Strang splitting scheme. In the right panel, we show the scaling of the error with the number of cells, for both Strang splitting schemes, computed on the test shown in the central panel.}
    \label{fig:DUSTYWAVE}
\end{figure*}
The next test is the standard DUSTYWAVE benchmark for the damping of a sound wave, originally presented in \cite{Laibe_2012} and extended to multiple grain sizes in \cite{Benitez-Llambay_2019}.  This is the first test in which both the hydrodynamical fluxes and the implicit drag terms are evolved simultaneously, making it especially important for assessing the performance of the Strang splitting scheme in coupling hydrodynamics with drag.

The gas and dust properties are initialized as small perturbations of the form
\begin{equation}
    \delta f = A \left[\text{Re}(\delta \hat{f})\cos{kx} - \text{Im}(\delta \hat{f})\sin{kx}\right],
\end{equation}
where the values for the real and imaginary parts of the perturbations are found in Table 2 of \cite{Benitez-Llambay_2019}. The amplitude of the perturbation is set to $A = 10^{-4}c_\text{s}$ for the velocities and $A = 10^{-4}\rho_0$ for the densities. The domain, $L=[0,1]$, is split into $N = 512$ evenly spaced cells, and the boundary conditions are set to be periodic.

\cite{Benitez-Llambay_2019} derived a general solution for any number of dust species, which we use to assess the validity of the numerical solution. Figure \ref{fig:DUSTYWAVE} shows the agreement between the numerical solution, obtained using the $\mathcal{D}_{\Delta t/2}\,\mathcal{H}_{\Delta t}\,\mathcal{D}_{\Delta t/2}$ Strang splitting scheme, and the analytical solution for both the single dust and the four dust-species cases. In addition, in the right panel of Fig. \ref{fig:DUSTYWAVE}, we show the scaling of the error with the number of cells, computed on the test with four dust species. As expected, both Strang splitting methods achieve second-order accuracy with increasing spatial resolution.

\begin{figure}[ht!]
    \centering
    \plotone{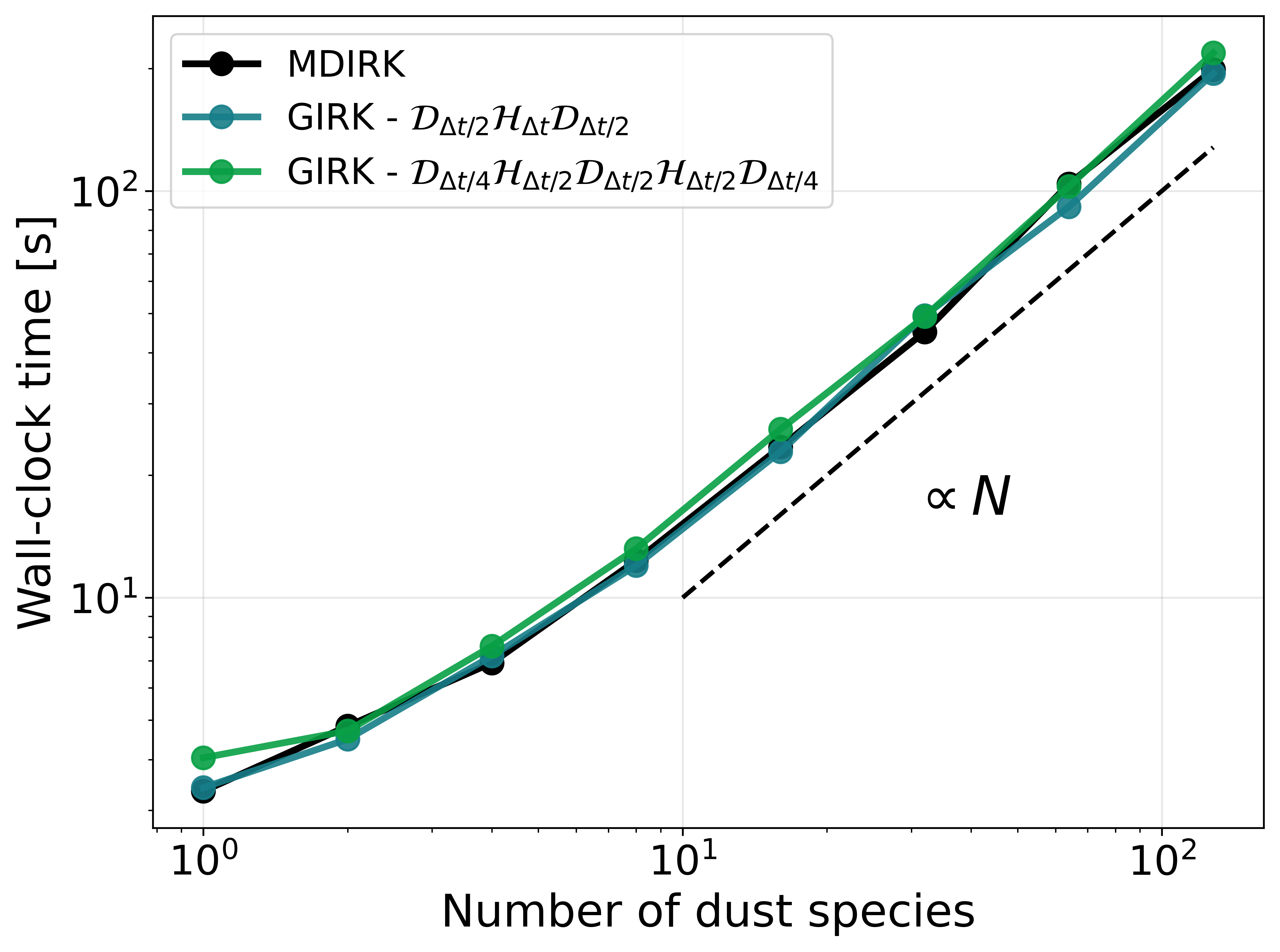}
    \caption{Wall-clock execution time for the DUSTYWAVE test for different number of dust species. By using the analytical solution and avoiding matrix inversion, all integrators scale linearly with the number of dust species.}
    \label{fig:timings}
\end{figure}

Like the MDIRK scheme, the GIRK integrator does not require matrix inversion and is therefore expected to maintain linear complexity as the number of dust species increases. To verify this, we conducted a modified version of the DUSTYWAVE test in which the original dust fluid is artificially split into $N_\text{d}$ identical fluids, all with the same stopping time but each representing a fraction of the original dust mass. This preserves the analytical solution while increasing the dimensionality of the drag coupling. We then performed simulations for $N_\text{d}$ ranging from 1 to 512 and measured wall-clock execution time. Figure~\ref{fig:timings} shows the timing results as a function of $N_\text{d}$, for the MDIRK integrator and both GIRK integrators proposed in this work. All methods show tightly clustered results around a linear trend, confirming the expected scaling behavior.

\subsection{DUSTYSHOCK}
\begin{figure*}[ht!]
    \centering
    \includegraphics[width=0.95\linewidth]{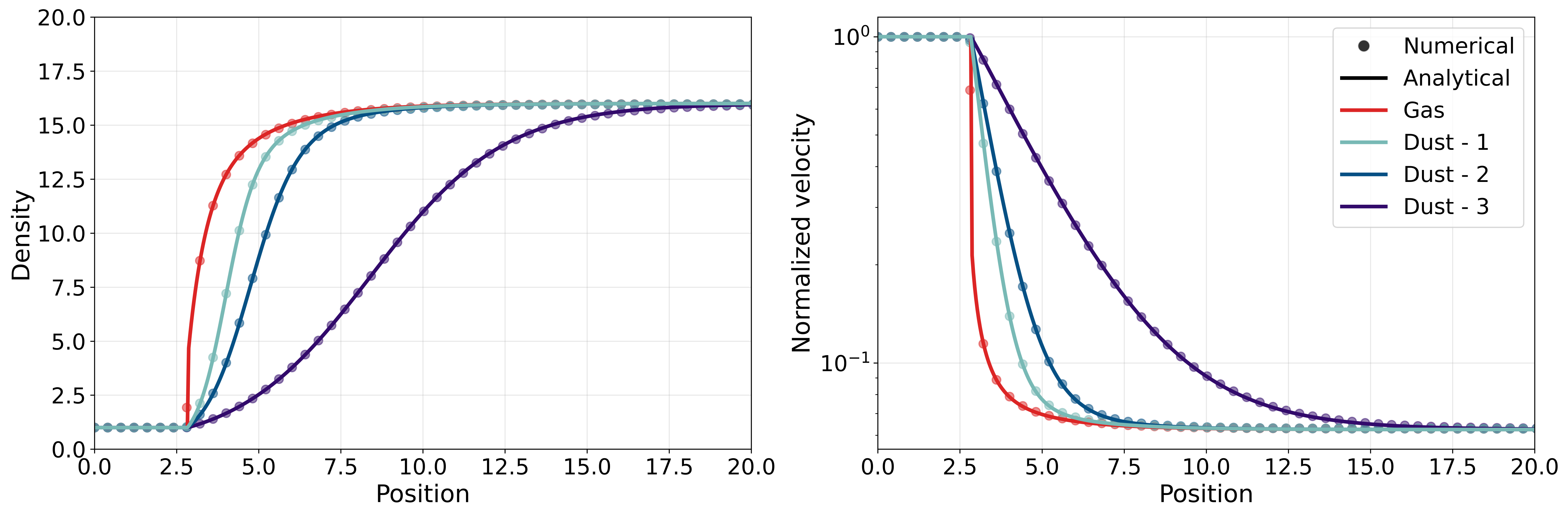}
    \caption{Density and normalized velocity of the gas and three dust species for the DUSTYSHOCK benchmark test. The figure shows a snapshot taken at $t = 500$ after the shock to ensure the system has had time to approach its steady state.}
    \label{fig:DUSTYSHOCK}
\end{figure*}
Unlike the linear wave dynamics examined in the DUSTYWAVE test, the DUSTYSHOCK benchmark evaluates the performance of drag integrators in a more challenging regime involving shocks and discontinuities. The original shock solution for a single dust species was presented in \cite{Lehmann_2018}, and later generalized to an arbitrary number of dust species by \cite{Benitez-Llambay_2019}. In this test, we adopt their analytical solution and initial conditions to evaluate the GIRK integrator, simulating a gas mixture with three dust species.

Because the analytical solution describes the post-shock steady state, we compare it with the numerical result only after the shock has fully propagated and the system has relaxed. Following the setup of \cite{Benitez-Llambay_2019}, we simulate a domain $x\in(0,40)$, discretized with 400 cells. The left state is imposed on the first 40 cells, while the remaining 360 cells represent the right state, creating a shock initially located at $x=4$. Transmissive (i.e., zero-gradient) boundary conditions are imposed. All fluids are initialized with the same left and right densities and velocities
\begin{equation}
    \rho_i^\text{L} = 1.0, \quad \rho_i^\text{R} = 16.0, \quad \omega_i^\text{L} = 1.0, \quad \omega_i^\text{R} = 0.0625,
\end{equation}
where $\omega_i = v_i/(\mathcal{M}c_\text{s})$ is the velocity normalized by the sound speed and the Mach number.\\ 
The three dust fluids differ only in drag coefficients, being $K_1 = 1.0$, $K_2 = 3.0$ and $K_3 = 5.0$, respectively. The sound speed is set to $c_\text{s} = 1.0$ and the Mach number to $\mathcal{M} = 2.0$.

Figure \ref{fig:DUSTYSHOCK} shows the density and normalized velocities of the gas and the three dust species at $t=500$ after the shock, alongside the numerical solution found by \cite{Benitez-Llambay_2019}. The integrator used was GIRK with the $\mathcal{D}_{\Delta t/2}\mathcal{H}_{\Delta t}\mathcal{D}_{\Delta t/2}$ Strang splitting scheme. The excellent agreement demonstrates that the GIRK integrator robustly handles drag even in the presence of strong shocks, while maintaining accurate coupling with the hydrodynamics.

\subsection{Steady-State 1D Shearing Box}
We conclude our test suite with the steady-state gas and dust drift problem in a one-dimensional shearing box. This configuration, first introduced by \cite{Krapp_2024}, represents the most complex setup in our suite, as it combines hydrodynamical fluxes, external forces, and non-trivial boundary conditions. The system includes centrifugal and Coriolis forces, as well as an additional term that mimics the radial pressure gradient in a protoplanetary disk, acting only on the gas momentum. Periodic boundary conditions are applied, with an extra source term in the $y$ component of the gas and dust momenta to reproduce the background azimuthal shear. The equilibrium solution for this problem was originally derived by \cite{Benitez-Llambay_2019} for multiple dust fluids, generalizing the classic single-fluid result of \cite{Nakagawa1986}. We adopt the equilibrium configuration described in Section 3.5.2 of \cite{Benitez-Llambay_2019} as the initial condition, allowing us to test the ability of our code to maintain this steady state over time.

The GIRK integrator was able, in both Strang splitting schemes, to maintain the initial conditions at equilibrium, with absolute variations with respect to the initial conditions smaller than $10^{-12}$, at a time of $t = 20$ in code units. This benchmark confirms that GIRK is able to resolve the dynamics of dust and gas even in complex environments, with external forces and non-standard boundary conditions.

\section{Discussion and Conclusion}
The integration of the drag term is a critical component in the dynamics of gas–dust mixtures involving multiple dust species. In recent years, several integrators have been developed to address this problem, with a particular emphasis on implicit schemes that avoid the restrictive timestep constraints imposed by small stopping times. While many of these approaches rely on matrix inversions, resulting in cubic computational complexity, the analytical solution introduced by \citet{Krapp_2020} and extended by \citet{Krapp_2024} provides a path to linear-time integration.

In this work, we formalized a systematic procedure to construct drag integration schemes following the analytical framework of \citet{Krapp_2020}, allowing us to evaluate their scaling and convergence properties. We first applied this procedure to the DIRK integrator presented in \citet{Krapp_2024}, coupled with a Strang splitting scheme, and found that it achieves second-order accuracy across most regimes. However, we also showed that DIRK fails to reach second-order convergence toward the equilibrium solution when external forces are applied and the timestep exceeds the maximum stopping time, i.e., for $\Delta t > t_{\text{s}}^{\text{max}}$. To overcome this limitation, we introduced a novel implicit drag integrator, GIRK (General Implicit Runge–Kutta), which generalizes the DIRK structure by lifting the diagonal constraint and introducing additional tunable parameters. This flexibility enables higher-order convergence while preserving linear complexity. Table~\ref{tab:convergence} summarizes the scaling properties of the various methods discussed in this paper, both in terms of accuracy of the drag integrator and convergence in the presence of an external forcing.
\begin{table}[h]
\centering
\caption{\label{tab:convergence} Accuracy and convergence orders for the DIRK and GIRK integrators across different timestep regimes.}
\begin{tabular}{l|c|c|c|c}
& \multicolumn{2}{c|}{Accuracy} & \multicolumn{2}{c}{Convergence} \\  
& $\Delta t < t_{s}^{\text{max}}$ & $\Delta t > t_{s}^{\text{max}}$ & $\Delta t < t_{s}^{\text{max}}$ & $\Delta t > t_{s}^{\text{max}}$ \\
\midrule
DIRK - $\mathcal{D}\mathcal{H}\mathcal{D}$   & II  & II  & II & I    \\
GIRK - $\mathcal{D}\mathcal{H}\mathcal{D}$   & III & II  & II & II   \\
GIRK - $\mathcal{D}\mathcal{H}\mathcal{D}\mathcal{H}\mathcal{D}$ & III & III & II & I / II \\
\bottomrule
\end{tabular}
\end{table}
Compared to the MDIRK integrator proposed in \citet{Krapp_2024}, GIRK offers a more versatile implementation into existing hydrodynamical codes through Strang splitting or similar operator-splitting techniques. Moreover, since it is derived using the same analytical foundation as DIRK and MDIRK, GIRK retains their key advantage: linear computational complexity with respect to the number of dust species. This makes it especially suitable for large-scale simulations involving processes like coagulation and fragmentation. Notably, GIRK relies on the same underlying approximation as DIRK: momentum exchange occurs only between the gas and each dust species, and not among dust species themselves. This assumption may break down in environments with highly concentrated dust clumps, which are uncommon in most astrophysical applications and would require specialized numerical treatments. To further evaluate GIRK’s performance, we plan to implement it in more complex hydrodynamical simulation codes, beyond the simple one-dimensional test code \texttt{pigpen} used in this work.\\

\begin{acknowledgements}
TB acknowledges funding by the Deutsche Forschungsgemeinschaft (DFG, German Research Foundation) under grant 325594231. GTP, TB, KD and BE acknowledge support by the Deutsche Forschungsgemeinschaft (DFG, German Research Foundation) under Germany's Excellence Strategy - EXC-2094 - 390783311.  GTP and KD acknowledge support by the COMPLEX project from the European Research Council (ERC) under the European Union's Horizon 2020 research and innovation program grant agreement ERC-2019-AdG 882679. GTP and TB acknowledge funding from the European Union under the European Union's Horizon Europe Research and Innovation Programme 101124282 (EARLYBIRD). Views and opinions expressed are, however, those of the authors only and do not necessarily reflect those of the European Union or the European Research Council. Neither the European Union nor the granting authority can be held responsible for them. MH acknowledges support from the DFG program ``Closing the Loop - Using Synthetic Observations of Simulated Star-forming Regions to Test Observational Properties" (DFG Project Number: 426714422).
\end{acknowledgements}
\appendix

\section{Variables for the Full Solution}
\label{sec:appendix_A}
We report in this appendix the definitions of the constants needed to compute the full analytical solution to the aerodynamic drag for the GIRK integrator to be used in Eq. \ref{eq:sol_beg} - \ref{eq:sol_end}.  
\begin{equation}
    \Lambda_i = \Bigl(1 + \alpha_i\Delta t\Bigl(\gamma_1 + \gamma_2 + \alpha_i\Delta t\,(\gamma_1\gamma_2 - \beta_1\beta_2)\Bigr)\Bigr)^{-1}, \quad \delta_{pi} = \bigl(1+\gamma_p\Delta t\alpha_i\bigr)^{-1},
\end{equation}
\begin{equation}
    A_p = \sum_{i=1}^{N_\text{d}} \alpha_i u_{di}\,\delta_{pi} - \beta_p\Delta t \sum_i^{N_\text{d}} \alpha_i^2 u_{di}\Bigl(1+\alpha_i\Delta t(\gamma_p - \beta_q)\Bigr)\delta_{pi}\,\Lambda_i,
\end{equation}
\begin{equation}
    B_p = \sum_i^{N_\text{d}} \alpha_i\epsilon_i\,\delta_{pi} - \beta_p \Delta t \sum_i^{N_\text{d}} \alpha_i^2 \epsilon_i\Bigl(1+\alpha_i\Delta t \,(\gamma_p - \beta_q)\Bigr)\delta_{pi}\,\Lambda_i,
\end{equation}
\begin{equation}
    C_p = \sum_i^{N_\text{d}} \alpha_i\epsilon_i\,\delta_{pi} - \Delta t \sum_i^{N_\text{d}} \alpha_i^2\epsilon_i(\gamma_q + \alpha_i \Delta t(\gamma_1\gamma_2 - \beta_1\beta_2))\delta_{pi}\Lambda_i,
\end{equation}
\begin{equation}
    D_p = 1 + \gamma_p\Delta t\sum_i^{N_\text{d}} \alpha_i\epsilon_i\,\delta_{pi} - \beta_1\beta_2\Delta t^2\sum_i^{N_\text{d}}\alpha_i^2\epsilon_i\delta_{pi}\Lambda.
\end{equation}
The index $i$ spans the $N_\text{d}$ dust grain sizes, whereas the index $p = 1,2$ differentiates between two similar sets of constants, while $q$ takes the opposite value with respect to $p$. 

\section{Stability analysis}
\label{sec:appendix_B}
In this appendix, we show the stability analysis for both GIRK integrators presented in Section \ref{sec:parameters_GIRK}. The \texttt{jupyter} notebook used to perform this stability analysis is freely available on \texttt{github}\footnote{\url{https://github.com/Giovanni-Tedeschi/GIRK_stability}}

The first step is to introduce the stability function for a general Runge-Kutta method
\begin{equation}
    R(\mu) = \frac{\text{det}(\mathbb{1} - \mu \mathbf{A} + \mu(\mathbf{e}\cdot \mathbf{b}))}{\text{det}(\mathbb{1} - \mu \mathbf{A})},
\end{equation}
where $\mathbf{e} = (1,1)$ and the $\mathbf{A}$ matrix and $\mathbf{b}$ vector are defined by the Runge-Kutta parameters. Following the choice we made throughout this paper
\begin{equation}
    \mathbf{A} = \begin{pmatrix}
        \gamma_1 & \beta_1 \\
        \beta_2 & \gamma_2
    \end{pmatrix},
    \quad 
    \mathbf{b} = (b, 1-b).
\end{equation}
The variable $\mu = \lambda\Delta t$ is the product between the timestep $\Delta t$ and the eigenvalues of the matrix defining the problem at hand; in this case, the matrix $\mathbf{M}$ in Eq. \ref{eq:mom_cons}.

The stability region associated to a Runge-Kutta method is the region in the $\mu$ plane where $|R(\mu)| < 1$; consequently, a Runge-Kutta method is considered A-stable if the stability region includes the entire left half-plane of $\mu$.

\begin{figure*}[h!]
    \centering
    \includegraphics[width=0.6\linewidth]{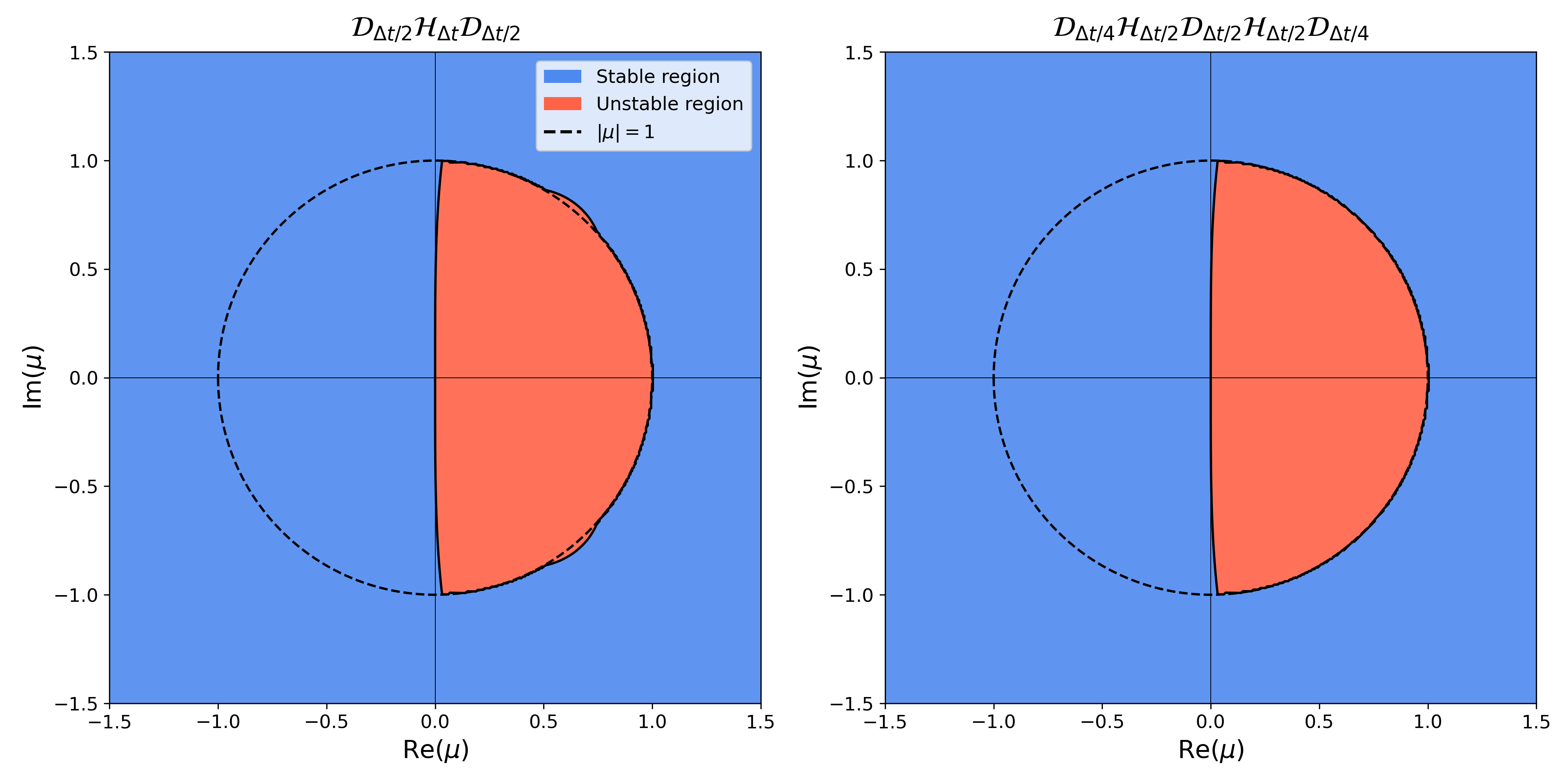}
    \caption{Stability regions of both Strang splitting schemes for the GIRK integrator. Being the left half-plane stable, both integrators are A-stable. The $|\mu| = 1$ circumference marks the boundary between the parameters chosen for $\Delta t < t_\text{s}^{\text{max}}$ and $\Delta t > t_\text{s}^{\text{max}}$.}
    \label{fig:stability}
\end{figure*}

In Figure \ref{fig:stability} we show the stability regions of both Strang splitting schemes introduced in Section \ref{sec:parameters_GIRK}. In this context, the only difference between the two methods lies in the choice of parameters, and thus of $\mathbf{A}$ and $\mathbf{b}$. In \cite{Krapp_2020}, the inverse of the maximum eigenvalue of $\mathbf{M}$ is derived to be bounded between the two largest stopping times of the system. Hence, the boundary between the two sets of parameters for each method, $\Delta t = t_\text{s}^{\text{max}}$, corresponds here to the circumference $|\mu| = 1$.
\\
As can be seen in Figure \ref{fig:stability}, both Strang splitting methods are A-stable, since their stability region contains the entire left half-plane. Although not shown in the Figure, we also checked that both methods are asymptotically stable, or L-stable, by checking that $R(\mu) \to 0$ as $\mu \to \infty$.

\bibliography{bibliography}{}
\bibliographystyle{aasjournalv7}



\end{document}